\documentclass[a4paper,aps,amsmath,amssymb,showpacs,10pt,twocolumn]{revtex4}
\usepackage{graphicx}
\usepackage{graphics}
\begin{document}
\draft
\title{Scaling and Renormalization Group in Replica Symmetry Breaking space:
Evidence for a simple analytical solution of the SK model at zero temperature}
\author{R. Oppermann$^{1}$ and D. Sherrington$^2$}
\address{$^1$ Institut f. Theoretische Physik, Universit\"at
W\"urzburg, Am Hubland, 97074 W\"urzburg, FRG}
\address{$^2$ Rudolf Peierls Centre for Theoretical Physics, University of Oxford,
1 Keble Road, Oxford OX1 3NP, UK}
\date{\today}
\pacs{75.10.Nr,75.30.Cr,71.23.An}
\begin{abstract}
Using numerical self-consistent solutions of a sequence of finite replica symmetry breakings
(RSB) and Wilson's renormalization group but with the number of RSB-steps playing a role of
decimation scales, we report evidence for a non-trivial $T\rightarrow0$-limit of the Parisi order
function $q(x)$ for the SK spin glass. Supported by scaling in RSB-space, the fixed point order function
is conjectured to be $q^*(a)=\frac{\sqrt{\pi}}{2}\frac{a}{\xi}\hspace{.1cm}erf(\frac{\xi}{a})$ on
$0\leq a\leq\infty$ where $\frac{x}{T}\rightarrow a$ at $T=0$ and $\xi\approx 1.13\pm0.01$.
$\xi$ plays the role of a correlation length in $a$-space.
$q^*(a)$ may be viewed as the solution of an effective $1D$ field theory.
\end{abstract}
\maketitle
It is now 30 years since, stimulated by the seminal work of Edwards
and Anderson \cite{EA}\footnote{DS thanks Prof Edwards for discussing his
discoveries with him in 1974, before their publication.},
one of the authors (DS) devised the canonical soluble mean-field spin glass model
\footnote{This model was exposed at an informal `brown-bag' seminar at Imperial College
in the spring of 1975.}
now known as the Sherrington-Kirkpatrick (or SK) model.
The first published report \cite{SK} already exhibited the `smoking gun'
which a few years later led Parisi
to his remarkable predictions of a hierarchy of replica symmetry breaking (RSB)\cite {Parisi3}
and its associated implications for macrostate complexity \cite{Parisi1} in disordered
frustrated systems.
These studies further stimulated new fields of investigation across physics, computer science,
biology,
econophysics and probability theory \footnote{For further descriptions of both the history and
the developments see the articles by Anderson, Sherrington, Parisi and M\'ezard in
\cite{GGS}, and also \cite{Guerra, Talagrand}.}. There has been great progress since then,
in studies of the original model and its many extensions; the concept of RSB is now
well-accepted for mean field models \footnote{Rigorous proofs are however just being
developed \cite{Guerra, Talagrand} and the corresponding situation for finite-range models
remains controversial.} and the behaviour of the Parisi order function is well studied
perturbatively near the transition temperature; there exist self-consistency equations
from which the Parisi function can be obtained in principle; and some features of RSB are
considered known at least semi-quantitatively down towards zero temperature.
However, here we demonstrate that there appear to be unappreciated subtleties
in the limit as zero temperature is approached. In particular, we exhibit evidence for a new
zero-temperature limiting Parisi order function with a new `correlation length' in
RSB-space and also unconventional behaviour of the $T=0$ local field distribution in the zero field limit.

\begin{figure}
\hspace{-.3cm}
\resizebox{.48\textwidth}{!}{%
\includegraphics{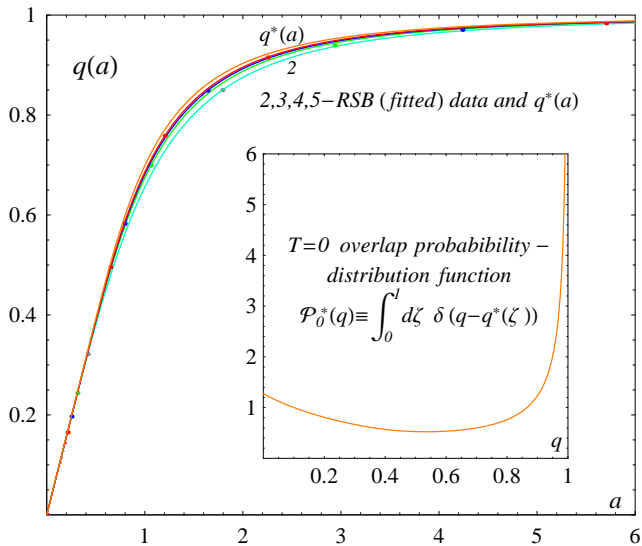}}
\caption{\label{fig:q2}Order parameter functions of the SK-model at zero temperature.
Shown as points are the numerical solutions of the finite RSB approximations
from 2nd (lowest curve) up to 5th order, together with fitting functions $q_{model}^{\alpha}(a)$ with
best-fit $\xi(\alpha)$ and $w(\alpha)$ and also the predicted limiting function $q^*(a)$
for $\xi=2/\sqrt{\pi}$ ($\xi=1.138$ is indistinguishable on the given scale).
Small dots, approximating the smallest $q_k(\alpha),a_k(\alpha)$ by extrapolating the
$\alpha$-dependence beyond the calculated $5th$ order, up to
$10th$ order, confirm the linear rise of $q^*(a)$.
The insert shows the overlap probability distribution function ${\cal P}^*(q)\equiv d\zeta(q)/dq$
derived from $q^*(a=\zeta/(1-\zeta))$.}
\end{figure}

\begin{figure}
\hspace{-.2cm}
\resizebox{.48\textwidth}{!}{%
\includegraphics{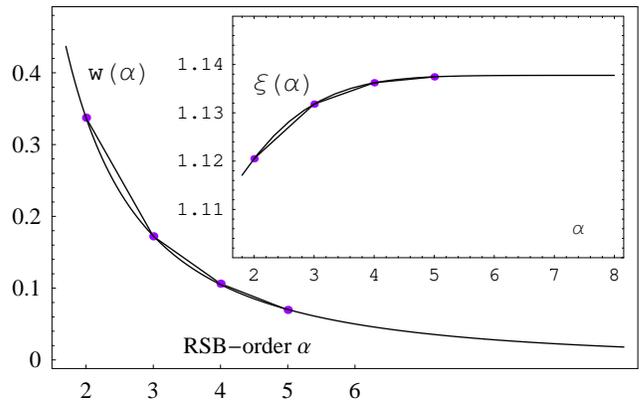}}
\caption{\label{fig:w-xi}Numerical renormalization group $RSB$-flow of the replica
correlation length $\xi(\alpha)$ (insert) and of the spreading parameter $w(\alpha)$
for the 2-parameter model (3-parameter model shows an equivalent flow)
with analytic fitting curves extrapolating to $lim_{\alpha\rightarrow\infty}\xi(\alpha)\approx 1.138$
and $lim_{\alpha\rightarrow\infty}w(\alpha)=0$ are included. Note that the evaluation of the points
shown requires a numerical accuracy far beyond that visible at the scale of Fig.\ref{fig:q2}}
\end{figure}

\begin{figure}
\hspace{-.5cm}
\resizebox{.48\textwidth}{!}{%
\includegraphics{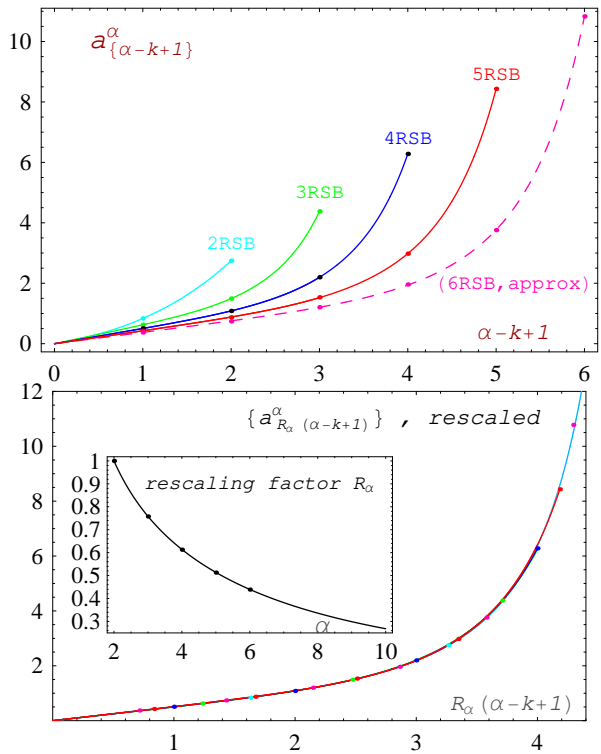}}
\caption{\label{fig:a-rescaling} Almost perfect mapping of $a_{k}$ parameters (top) onto a single curve
(below) by rescaling with $\alpha-k+1\rightarrow R_{\alpha}(\alpha-k+1)$ in RSB space;
$R_{\alpha}$ is shown in insert as a function of RSB-order. A similar
transformation exists for the $q$ parameters.}
\end{figure}

\begin{figure}
\resizebox{.48\textwidth}{!}{%
\includegraphics{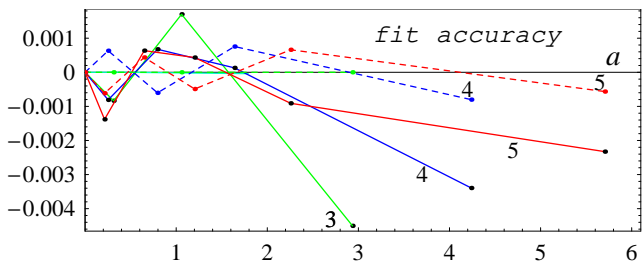}}
\caption{\label{fig:q2b}
Quality of fits illustrated by the distance between numerical data and fit curves for the
2- and 3-parameter models (2pm and 3pm, dashed). Labels indicate RSB order.
Misfits decrease from $O(10^{-3})$ for the 2pm to $O(10^{-4})$ (3pm) and improve with RSB order for both,
while the numerical inaccuracy of $q$- and $a$-parameters is $O(10^{-5})$ and $O(10^{-4})$ resp.
}
\end{figure}

The models we consider are the usual SK one \cite{SK}
\begin{equation}
{\cal H}=\sum_{i<j} J_{ij} \sigma_i \sigma_j;\hspace{.2cm} \sigma=\pm1
\end{equation}
and the Fermionic Ising Spin Glass (FISG) \cite{Oppermann, OS2003}
\begin{equation}
{\cal H}=\sum_{i,j}J_{ij}\sigma^z_i \sigma^z_j -\mu\sum_i (n_{i\uparrow}+n_{i\downarrow});\hspace{.2cm} n_{i,s}=0,1
\end{equation}
in the half-filling limit, where the spin variables are given by
$\sigma^z=a^{\dagger}_{\uparrow}a_{\uparrow}-a^{\dagger}_{\downarrow}a_{\downarrow}$ and
$a^{\dagger}_{s}$ and $a_{s}$ are fermion creation and annihilation operators.
\footnote{The thermodynamics \cite{Oppermann} and the spectral density
functions \cite{Isaac}
of the {$\mu = 0$} FISG  map onto
properties of the SK model; they have an identical Parisi $q(x)$
and the one particle density of states $\rho(\epsilon)$ of the FIRSG  is given by the SK
$p(h)$ with $h=\epsilon$}.
The $J_{ij}$ are quenched disordered and frustrated interactions independently
distributed over all pairs of sites as
\begin{equation}
P({J})=exp(-NJ_{ij}^2/(2 J^2))/(\sqrt{2\pi J^2 /N}).
\end{equation}
We measure temperature in units of $J$ and for notational convenience set the energy scale $J=1$.
We shall consider its properties in replica theory and assume the Parisi ansatz as an infinite
sequence of replica symmetry breaking (RSB) steps for an order function $q(x)$ defined
on $0\leq x\leq 1$ \cite{Parisi3, MPV}. Denoting the steps as $x_k$ it is believed
\cite{PAT, Crisanti2002} that in the limit $T\rightarrow 0$ and small $x \leq O(T)$
the steps scale with $T$ and the natural variables are $T$-normalized steps $a_k$ given by
\begin{equation}
a_k\equiv lim_{T\rightarrow0}\hspace{.2cm}(J/T)x_k(T),
\label{def:anorm}
\end{equation}
which become selfconsistently distributed on the interval $0\leq a\leq\infty$.
We focus on the $T=0$ limit and express the free energy functional of the Parisi
scheme at $\alpha$-RSB in terms of the variational parameter set
{$\{q_{k+1},a_k ;k=1...{\alpha}\}$};
we also have constrained $\{q_1=1, q_{\alpha+2}=0\}$
\footnote{Details of the expression for the free energy and results, as well
as a discussion of the logical steps guiding our analysis, are deferred to a longer paper \cite{OS2005}.}.
The extremization has been performed for $\alpha=1...5$ to an accuracy of 4 decimal places.
The resulting values of $q_k,a_k$ are shown in Fig.\ref{fig:q2} for
$\alpha =$ 2- to 5-RSB, together with fitting functions of the form
\begin{equation}
q_{model}^{\alpha}(a)=\frac{\sqrt{\pi}}{2}\hspace{.2cm}\frac{a}{\xi(\alpha)}
\hspace{.1cm}erf{\left[\xi(\alpha)/\sqrt{f_{\alpha}(a)}\right]}.
\label{eq:q-model}
\end{equation}
where $f_{\alpha}(a)$ is chosen as $(a^2+w(\alpha))$ in a $2$-parameter model and as
$(a^2+c(\alpha)a+w(\alpha))$ in a refined $3$-parameter model.
Both choices derive from the ansatz
of a Gaussian distribution of error-functions which (in contrast to a single error function) fit well the
numerical data.
The $\xi(\alpha)$, $w(\alpha)$ and $c(\alpha)$ are viewed as parameters in an RG decimation-type flow space
with the $\alpha$ playing the role of the degree of decimation.
Fig.\ref{fig:w-xi} shows these flows together with
analytic fit functions which extrapolate as $\alpha\rightarrow\infty$
to  $lim_{\alpha\rightarrow\infty}\xi(\alpha)\approx 1.14$ and $lim_{\alpha\rightarrow\infty}w(\alpha)=0$
(and for the 3-parameter case also $lim_{\alpha\rightarrow\infty}c(\alpha)=0$). Fig. \ref{fig:q2b}
shows the quality of the fits.
Moreover, if the indices $k$ are considered as potentially real continuous variables,
single parameter rescalings ${R_{\alpha}k}$ and ${T_{\alpha}k}$ are found to map
$a$ and $q$ respectively almost onto single curves, as Fig.\ref{fig:a-rescaling} shows,
with small decreasing deviations as $\alpha$ increases. This scaling behaviour in RSB-space of
the k-parametrized order parameter function supports a single parameter limiting form. Hence we
conjecture that the infinite RSB limiting form of the $T\rightarrow0$ Parisi order function is
\begin{equation}
q^*(a)=\frac{\sqrt{\pi}}{2}\hspace{.1cm}\frac{a}{\xi}\hspace{.1cm}
erf{\left(\frac{\xi}{a}\right)}.
\label{eq:qstar}
\end{equation}
Thus we predict that there is a finite intrinsic correlation length $\xi$ in RSB-space.
As noted above, our numerical results fit $\xi{(\alpha=\infty)}\approx 1.138$.
Assuming low T PaT-scaling\cite{PAT,VPT,Crisanti2002} $q(x,T)=q(x/T)$ up to $x=x_{max}(T)$
and $q(T)\sim1-O(T^2)$ for $x>x_{max}$, fitting the Gibbs susceptibility
$\chi(0)
=\lim_{T\rightarrow0}\beta\int_0^1 dx(1-q(x,T))
=\int_0^{\infty}da(1-q(a))=\frac12 \sqrt{\pi}\hspace{.1cm}\xi$
to its expected value $\chi(T=0)=1$ in the $\alpha=\infty$-limit requires
$\xi=2/\sqrt{\pi}\hspace{.1cm}\approx 1.128$.
Applying the same scaling assumption to the internal energy $U(T)=-\frac12 \beta\int_0^1 dx(1-q^2(x,T))$
we obtain $\xi\approx 1.1255$ by equating $U(0)$ and the $T=0$-limit of the free energy
\footnote{The slight differences in these figures for $\xi$ probably reflect on an imprecision
of the PaT-hypothesis\cite{Crisanti2002}}.
The smallness of this length suggests why already low orders ($\alpha>2$) of RSB produce good results.
Fig.\ref{fig:q2} compares our prediction with the finite-RSB results.

\begin{figure}
\hspace{-.2cm}
\resizebox{.48\textwidth}{!}{%
\includegraphics{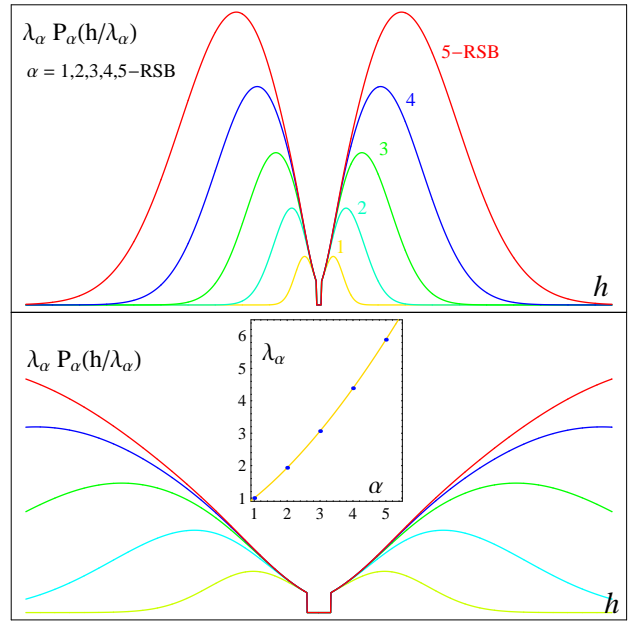}}
\caption{Rescaled field distributions $\tilde{P}_{\alpha}(h)$ (top: full scale, RSB orders labelled,
bottom: gap region zoomed). The Insert shows
the rescaling parameter $\lambda_\alpha$ as a function of RSB order.}
\label{fig:pscaled}
\end{figure}

The overlap probability distribution (OPDF) for finite temperatures is given by\cite{MPV}
${\cal P}(q)\equiv\int_0^1 dx\hspace{.1cm}\delta(q-q(x,T))=dx/dq$,
where $x(q,T)$ is the inverse of the original Parisi function $q(x,T)$.
The formulation for $q^*(a)$ which we have determined can be considered valid only for
low $x$ scaling as $T$ but suffices to give ${\cal P}(q)$ for $q < O(1)$.
Rescaling $a$ to a unit interval via $a\rightarrow\zeta=a/(1+a)$, the contribution of
$q^*(\zeta)$ to ${\cal P}(q)$ is $T(1-\zeta^*(q))^{-2}{\cal P}_0^*(q)$ where
${\cal P}_0^*(q)=\int_0^1d\zeta\hspace{.1cm} \delta(q-q^*(\zeta))=d\zeta^*(q)/dq$ is the $T=0$ analog OPDF.
This function obeys a non-algebraic equation yet allows one to extract an analytical result
for the small-$q$ gap and a $q=1$ divergence as
\begin{equation}
{\cal P}_0^*(q=0)=4/\pi,\hspace{.2cm}{\cal P}_0^*(q\approx 1)=\frac{\sqrt{3}}{4\sqrt{\pi}}
\frac{1}{\sqrt{1-q}}
\end{equation}
together with  a minimum at $q_{min}\approx 0.535$. Note, however, that
the linearity of dependence of $q(x)$ on $x$ only as $a=x/T$ is strictly valid only
for $a \leq O(1)$ and there is predicted a plateau of constant q above $x_{max}$ of order
0.5. This yields a delta function in ${\cal P}(q)$ close to $q=1$ and approaching $q = 1$
as $T \rightarrow 0$.
\footnote{Our results are also qualitatively and semi-quantitatively comparable with results of a
study of $q(\beta x)$ using a numerical analysis of high order perturbation theory
in the reduced temperature $(T_c - T)/T_c$, where $T_c$ is the spin glass transition
temperature \cite{Crisanti2002}. Both these methods can be considered limited but in
complementary, though possibly related, ways; our analysis is explicitly for zero temperature
but extrapolates from a finite number of steps of replica symmetry breaking, whereas the perturbation
method is based on full replica symmetry breaking but is limited by the perturbation being around $T_c$.}
\footnote{Note that the negativity of the slope of $P_0^*(q)$ at and near $q=0$
is removed by the prefactor $(1-\zeta^*(q))^{-2}$, which yields a flat $P(q)/T$ at $q=0$ in agreement with
Ref.\onlinecite{Crisanti2002}}.
One might note that the original finite-step Parisi equations have been
considered analytically in the continuum limit
as the number of steps go to infinity, to yield $q(x)$ as a solution of a set of coupled
partial differential equations with appropriate boundary conditions \cite {Parisi3,SD}
and in principle this could be continued to zero temperature. Unfortunately, however,
the practice is more difficult
and so far not effected analytically. Numerical solution is again possible in principle but,
since the boundary condition is at the transition temperature,
it presents accuracy problems to reach
comparison with the results of finite RSB reported here. Hence one is led instead to use
the extrapolation reported above to suggest an underlying `field theory'. To this end we note
that $q^*(a)$ is a solution of the differential equation
\begin{equation}
\left[\frac{1}{2} \hspace{.1cm}\frac{d^2}{d\tilde{a}^2}-\frac{1}{\tilde{a}^3}\frac{d}{d\tilde{a}}
+\frac{1}{\tilde{a}^4}\right]Q(\tilde{a})=0,
\label{eq:a-diff.eq.}
\end{equation}
where $\tilde{a}\equiv{a}/{\xi}$. It can also be conveniently expressed as a simple
generalized equation of motion in $a$-space through a substitution, to yield (dropping the tilde on a)
\begin{equation}
\frac{d}{da}\psi(a)=-\frac{\delta {\cal H(\psi)}}{\delta\psi(a)},\hspace{.3cm}
\psi(a)=e^{1/a^2}\frac{d\hspace{.1cm}log(Q(a))}{da}
\end{equation}
with a cubic `Hamiltonian'
\footnote{Contributions from higher orders of $exp(-1/a^2)\psi(a)$
cannot be distinguished neither for small nor for large $a$ due to numerical accuracy limitations.}
\begin{equation}
{\cal H}(\psi)=\int da \left(2a^{-4}e^{1/a^2}\psi(a)+\frac13 e^{-1/a^2}\psi(a)^3\right).
\label{eq:eqm}
\end{equation}
Now we turn to the local field distribution
\begin{equation}
P(h)=N^{-1}\sum_i\langle{\delta(h-\sum_j{J_{ij}}\sigma_j)}\rangle,
\end{equation}
again in the limit $T\rightarrow0$. The results for $P(h)$ of 0-5 RSB show that for any
finite RSB there is a gap around $h=0$ which reduces with RSB order but can be made
invariant by rescaling the field and (inversely) the probability $P$ with an
$\alpha$-dependent factor $\lambda_\alpha$ according to
\begin{equation}
\tilde{P}_{\alpha}(h)=\lambda_{\alpha}P_{\alpha +1}(h/{\lambda_{\alpha}})
\end{equation}
with $\lambda_{\alpha}$ chosen to yield the same gap for each $\alpha$. This
results in a set of $\tilde{P}_{\alpha}(h)$ with good overlap with one another
as $h$ tends towards the gap edge; this is illustrated in Fig. \ref{fig:pscaled}.
Extrapolating to $\infty$-RSB yields the $h\rightarrow0$ fixed point equation
\begin{equation}
{P}^*(h)=\lambda_{\infty} {P}^*(h/\lambda_\infty).
\end{equation}
The numerical results for the rescaling parameter $\lambda$ show an almost
linear increase with $\alpha$, confirming that in the $\infty$-RSB limit $P^*(0)=0$.

\begin{figure}
\resizebox{.48\textwidth}{!}{%
\includegraphics{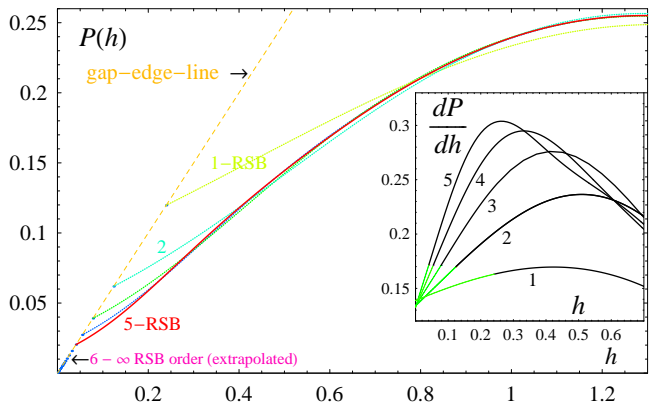}}
\caption{\label{fig:smallh}Field distribution $P(h)$ close to $h=0$; the lines end at the gaps in the
distributions. The inset shows $dP(h)/dh$ with solid lines corresponding to regions of non-zero
$P(h)$ and dotted lines showing their linear extrapolation to $h=0$.}
\end{figure}

A crude look at the limiting $P(h)$ for $h$ less than of order 0.5 shows a linear slope
of $P(h)$ of order 0.3, which has been suggested as the correct limit for many years
\cite{TAP,KS,PP,SD,Thomsen}. However a close look at the approach to the gap edge
suggests that the situation is more subtle. This is illustrated in
Fig. \ref{fig:smallh} where it is seen that the limit of $h\rightarrow0$ has a much smaller
slope of approximately 0.17 with a slope of order 0.3 taking over only at slightly
larger $h$. The
figure insert for $dP(h)/dh$ amplifies this observation as (i) the limit of the gap-edge
values of $dP(h)/dh$ at 0.17 and (ii) a peak in $dP(h)/dh$
around 0.3 at slightly higher $h$
\footnote{It is presumably this peak which is responsible for the larger slope observed in earlier studies.
It is however impossible at the RSB order we have studied to be sure whether the peak in $P'(h)$
around 0.3 saturates at a finite $h$ or tends towards zero field, while equally, from Fig \ref{fig:smallh},
it is difficult to believe that the slope at the gap edge does not tend to the lower value.}.

In summary, via scaling and a new renormalization group in RSB space,
for the first time a surprisingly simple analytical model solution $q^*(a)$
and a corresponding solvable one-dimensional effective field theory are proposed for
spin glasses in the low temperature limit, inviting the design of approximations
which may cope with more complicated problems of related but non-solvable natures.
Similar considerations have raised novel issues concerning the pseudogap in
the distribution of local fields and densities of states.

This work has been supported by DFG through grant Op28/5-2 and D5 of the SFB410,
EPSRC through grant GR/R83712/01, and ESF through programme SPHINX.

\end{document}